\documentclass[fleqn,usenatbib]{mnras}

\usepackage{newtxtext,newtxmath}
\usepackage[T1]{fontenc}

\DeclareRobustCommand{\VAN}[3]{#2}
\let\VANthebibliography\thebibliography
\def\thebibliography{\DeclareRobustCommand{\VAN}[3]{##3}\VANthebibliography}

\usepackage{graphicx}	% Including figure files
\usepackage{amsmath}	% Advanced maths commands

\usepackage[dvipsnames]{xcolor}
\usepackage{booktabs} %table 3 
\usepackage{multirow} % table 3
\usepackage{subcaption}
\title[Evolution of a $ 9\rm M_\odot$ progenitor]{Evolution and formation of ultramassive white dwarf stars: The case for a $ 9\rm M_\odot$ progenitor}

\author[A. S. R. Antonini et al.]{
Ana S. R. Antonini,$^{1}$\thanks{E-mail: anacarolina.srantonini@gmail.com}
Alejandra. D. Romero,$^{1}$
S. O. Kepler$^{1}$
\\
$^{1}$Instituto de Física, Universidade Federal do Rio Grande do Sul, 91501-970 Porto Alegre, RS, Brasil\\}

\date{Accepted 2026 March 04. Received 2026 February 23; in original form 2025 November 10}

\pubyear{\the\year{}}

\begin{document}
\label{firstpage}
\pagerange{\pageref{firstpage}--\pageref{lastpage}}
\maketitle

\begin{abstract}
We study the full evolution of a 1.313\,$ \rm M_\odot$ white dwarf star that descended from a 9\,$ \rm M_\odot$ main-sequence progenitor with an initial metallicity of Z=0.02. Using MESA r24.08.01, we calculate its entire evolution from pre-ZAMS to the WD cooling curve, including both the evolution through 139 thermal pulses and the post-AGB phase. The resulting remnant is an ultramassive H-deficient WD, for which the composition, in mass fraction, is 47.7\% $^{16}$O, 39.7\% $^{20}$Ne, 4.2\% $^{24}$Mg, 3.3\% $^{23}$Na and 0.386\% $^{12}\rm C$ -- corresponding to a total mass of $5 \times 10^{-3} \rm\, M_\odot$ of C --, surrounded by a  $1.5 \times 10^{-5}\rm M_\odot$ He layer. We also investigate the effects of fully suppressing the TP-SAGB stage by adopting a high mass-loss rate only after the second dredge-up, and find only minor differences in the final mass and composition. In addition, we calculate models with and without phase separation during the WD stage, estimating a cooling delay of only 16~Myr.   %compare the effects of phase separation during the WD stage, estimating a cooling delay of only 16~Myr. 
 This is the first ultramassive white dwarf sequence for which both the TP-SAGB and post-AGB stages are calculated and, to our knowledge, the most massive WD model from complete evolution for which cooling times and detailed abundance profiles are published. 
\end{abstract}

\begin{keywords}
white dwarf stars -- stellar evolution -- stellar modeling
\end{keywords}

%%%%%%%%%%%%%%%%%%%%%%%%%%%%%%%%%%%%%%%%%%%%%%%%%%

%%%%%%%%%%%%%%%%% BODY OF PAPER %%%%%%%%%%%%%%%%%%

\section{Introduction} \label{sec:intro} 
White dwarf (WD) stars are the final stage of evolution for low- and intermediate-mass stars, which account for more than 95\% of the stars in the Galaxy. Their relatively simple and slow evolution, primarily driven by cooling, makes them excellent cosmic clocks and testbeds for new or high-density physics \citep{Isern2008_axionWD,GarciaBerro2014_wd47tuc,Campo2016_wd_gcs}. In addition, massive and ultramassive white dwarfs may become type Ia supernovae, which are used as standard candles to detect the accelerated expansion of the Universe, if mass is accreted from a companion or through the merger of two white dwarfs.
 % being widely used for dating nearby stellar populations. 
%Their numerous classes of progenitors translate into a wide mass distribution. And 

For the population of white dwarf stars, the stellar mass spans a wide range, from $\sim 0.2$ to $\sim 1.3\, \rm M_\odot$, with the maximum near $0.6\,\rm M_\odot$, and secondary maxima at $\sim0.39\,\rm M_\odot$ and $\sim0.8\,\rm M_\odot$ representing a low-mass component and a massive component \citep{Kepler2007,Kilic100pc,Obrien40pc}.

White dwarfs with masses above $\sim 0.3 \rm M_{\odot}$ should have C/O cores, while those with stellar masses above 1.05\,$\rm M_\odot$, called Ultramassive White Dwarfs (UMWD), are expected to harbor ONe cores as a result of carbon burning in the (Super) Asymptotic Giant Branch [(S)AGB] \citep{garcia-berro+1997}. 

Unlike the theoretical maximum mass allowed for a non-magnetic, non-rotating white dwarf,  which is a well-established limit known as the Chandrasekhar mass ($\rm M_{Ch}\approx 1.44\,\rm M_\odot$), the maximum mass for a WD progenitor remains much more contentious. Different theoretical works yield different values, which range between $\sim$7 and 12\,$\rm M_\odot$ \citep{siess2010,doherty+2015,lauffer}. This is not merely an effect of metallicity, as different works that adopt the same initial mass and metallicity still find different initial-to-final mass relations (IFMR). Instead, it appears to be the combined effect of adopting different physical prescriptions in the model calculation. The most important contributors to these discrepancies are, by far, the treatments of convection and mass loss. 

The treatment of convective boundaries in the core, especially during core H and He burning (CHB and CHeB), strongly impacts the mass of the He-free core ($\mathrm{\rm M_{CO}}$), changing not only the final mass of a stellar remnant but, in some cases, even its very nature.  For example, the addition of even modest amounts of core overshooting during CHB and CHeB can lower by $\sim 2-3\,\rm M_\odot$ the initial mass needed for a star to undergo degenerate carbon-burning \citep{Eldrige+Tout2004,siess2010,timmes+2015}. The different computations, physical ingredients, and parameters --- especially those concerning the treatment of convection and overshooting --- are poorly constrained by observations and contribute to uncertainties in the theoretical IFMRs. 

Regarding the role of stellar winds, the existing uncertainties in mass-loss formulations during the AGB \citep{Karakas2014_review,Decin2021_mass_lossrev,Matthews2024} become even larger for SAGB stars. There are several reasons for this, the main being that: 1 - stars with higher masses lose a larger fraction of their initial mass than their lower-mass counterparts; 2- SAGB stars lose only a very small portion of their envelope during the prior stages of evolution, meaning most of their mass loss occurs during the AGB.  Additionally,  the mechanism driving winds in SAGB stars is still unclear \citep{Poelarends_2008,Hofner18_mdot,Beasor2020,Cheng_2024_mass_loss_mech},  and mass-loss rates during this stage are, in general, poorly constrained by observations, varying by over three orders of magnitude \citep{vanLoon1999,Hofner18_mdot,Matthews2024}.

Another, albeit smaller, factor that complicates establishing a more accurate maximum mass limit for UMWD progenitors is their evolution into SAGB stars. They are expected to ignite carbon under degenerate conditions and undergo several episodes of carbon flashes followed by carbon flame propagation, depending on the location of the first ignition inside the star. Additionally, these stars will undergo several tens to hundreds of thermal pulses during the AGB, making them both challenging and computationally expensive to model.  After several works dedicated specifically to it \citep{Timmes1994_Cflame,Siess2006,Denissenkov_2013,timmes+2015}, modeling the first of these two stages is now relatively straightforward.  However, dealing with the end of the TP-SAGB remains an issue. There are no solutions to the instabilities that arise at this stage, causing computations to stall before the star enters the post-AGB phase  \citep{HRI,Lau+2012,Rees+Izzard}. Hydrogen Recombination Instability (HRI), which is now reasonably managed, arises when the envelope becomes extended and cool enough to allow hydrogen to recombine rapidly. The energy generated causes the envelope to further expand and cool, eventually leading to a runaway expansion that hydrostatic 1D codes cannot track. The Iron Peak Instability (FeI), however, remains an undefeated foe for TP-SAGB modelers. It appears when  Iron group elements create an opacity peak at the base of the convective envelope, causing a buildup of energy in the region. This increases local radiative pressure, which becomes dominant over the gas pressure. This small  $\mathrm{P_{gas}/P}$ ratio
 progressively decreases, forcing calculations to adopt smaller and smaller timesteps and eventually to stop converging altogether once the ratio approaches zero. As a result of these instabilities, there are no Ultramassive WD sequences in the literature for which both the TP-AGB and post-AGB stages were computed.   

In the past, three distinct approaches have been used to develop UMWD models. The first considers a model at the top of the cooling curve as the initial model, where the inner structure of the core, mainly the chemical composition, is taken from a SAGB model before its envelope is stripped \citep[e.g.][]{Corsico+2004,camisassa+19,Althaus2021}. The second also starts with a model at high effective temperatures, but it assumes that the composition is not mass-dependent and either rescales existing WD models to higher masses or creates new UMWD models with a fixed composition \citep[e.g.][]{Benvenuto1999,Schwab_2021,Althaus2022}. The third approach involves fully bypassing the thermal pulses by adopting extreme mass-loss rates in the AGB phase early on \citep[e.g.][]{lauffer,Castro-TapiaPS}.   

In this work, we propose an intermediate approach. We calculate a significant fraction of the thermal pulses the star is expected to undergo and force an exit from the TP-AGB before those instabilities can halt our calculations, thus allowing the model to undergo post-AGB evolution and finally reach the WD cooling curve. 
This technique has previously been used by \citet{DeGeronimo+18-TPs} to investigate the impact of thermal pulses on the properties of DAVs, and is a somewhat common approach to modeling the previous evolution of CO WDs. We employ the MESA code (\citep{mesa1,mesa2,mesa3,mesa4,mesa5,mesa6} for our calculations and choose a 9\,$\rm M_\odot$ Z=0.02 sequence that generates a 1.313\,$\rm M_\odot$ WD. With this choice, we address in one go the lack of fully evolutionary white dwarf models near the Chandrasekhar mass limit and the lack of UMWD models that account for previous evolution, both in the TP-AGB and post-AGB. 
\par This paper is organized as follows. We describe the input physics adopted in our calculations at each evolutionary stage in Section~\ref{sec:models}. Results and comparisons are presented in Section~\ref{sec:results}. In Section~\ref{sec:conc}, we conclude with our final remarks.

\section{Stellar Modeling} \label{sec:models}
We calculate the complete evolution --- from pre-ZAMS to the WD cooling track --- of a sequence with initial mass of $9\,\rm M_\odot$ and metallicity Z=0.02, using the unidimensional stellar evolution code MESA [Modules for Experiments in Stellar Astrophysics \citep{mesa1,mesa2,mesa3,mesa4,mesa5,mesa6}]
in version r24.08.1. These parameters have been chosen due to the availability of works for comparison \citep[e.g.][]{lauffer,camisassa+19}. \par
We adopt the nuclear reaction network \verb|mesa45.net|, which contains 45 isotopes and 367 reactions. This is motivated by the fact that the commonly adopted MESA network \verb|sagb_NeNa_MgAl.net| results in non-negligible amounts of free neutrons in the remnants, as it accounts for the production of neutrons via the $\mathrm{^{13}C(\alpha,n)^{16}O}$ reaction, but not for their capture. Abundance fractions are taken from \citet{Asplund+2009}, and our reaction rates are mainly from JINA \citep{jina} and NACRE \citep{nacre}, with a small number of rates from CF88 \citep{CF88}. Supplementary tabulated weak rates are from \citet{weak1},  \citet{weak2}, and \citet{weak3}. The thermal neutrino loss rates are from \citet{itoh96}, and the screening prescription is from \citet{screening}. 

We adopt the OPLIB opacity tables \citep{Colgan+2016,Farag+2024}, complemented by \citet{Ferguson+2005} at low temperatures, and also consider Type 2 opacity tables to account for C and O enrichment. %after the start of CHeB.
Electron conduction opacities are from \citet{Cassisi+07} complemented by those of \citet{Blouin2020_conductive_op} in the regime of moderate coupling and degeneracy, for both H and He. We do not consider rotation.
Below, we describe in detail the parameters used for each stage in the evolution.

\subsection{Pre-WD evolution} \label{subsec:prewd}

During the stages preceding the white dwarf cooling curve, the atmosphere boundary is calculated using the gray Eddington t(tau) relation with uniform opacity. We consider the Mixing Length Theory (MLT) \citep{MLT} for treating convection, using the \cite{Henyey65} formalism, which allows convective efficiency to vary with opacity, with $\alpha_{\rm MLT} = 2$, which is a commonly adopted value and
within the range of general calibrations \citep{Noels91}. We detail the specific convection and mass-loss prescriptions, as well as resolution controls adopted at each evolutionary stage, in the following subsections. 

\subsubsection{Core Hydrogen and Core Helium Burning}
In our treatment of convection from ZAMS until the end of CHeB, we consider the Ledoux criterion, adopting semiconvection in the \citet{Langer85} scheme with  $\alpha_{\rm semi} =0.05$ and thermohaline mixing in the \citet{kip+80} method, with $\alpha_{\rm th} = 1$. We further extend the formal convective boundaries by adopting exponential overshooting at the top of the H and He cores and shells, with $f=0.01$ and $f_0=0.005$ \citep{herwig2000}. In addition, we adopt the predictive mixing scheme present in MESA \citep{mesa4} to prevent splitting of the core convective region and the occurrence of breathing pulses. 
\par Mass loss is treated using the formula from \citet{reimers}, with $\eta_R = 0.3$, which is an arbitrary value as, at this initial mass, very little mass will be lost before the TP-(S)AGB stage. We account for element diffusion, adopting the coefficients from \citet{Stanton+Murillo}.

\subsubsection{Early AGB and Carbon Burning}
As bottom overshooting and thermohaline mixing have been found to dilute the carbon flame, giving origin to hybrid ONeC cores \citep{Denissenkov_2013,timmes+2015},
we turn off thermohaline mixing and restrict overshooting to occur only at the upper boundaries of H and He burning shells and non-burning convective zones. We maintain our choices of semiconvection and account for diffusion only in timesteps longer than 5 years.   
Mass loss is treated with the \citet{bloecker} formulation, with $\eta_B=0.1$, which is an intermediate value chosen due to a lack of strong constraints on SAGB mass loss, and a maximum allowed mass-loss rate ($\dot{\rm m}_{\rm max}$) of $10^{-6}\rm M_\odot/\mathrm{\rm yr}$ to prevent very high mass-loss rates before the onset of thermal pulses.

Carbon burning occurs under degenerate conditions for SAGB stars, making this process very sensitive to temporal and spatial resolution \citep{Siess2006,timmes+2015}. Thus, we apply extra meshing in the CO core and further increase meshing in $^{12}$C--$^{12}$C-burning regions. This is done in the MESA inlist using the controls \verb|mesh_logX_species| and \verb|mesh_dlog_cc_dlogP_extra|. Additionally, we also require changes in $\log$L\_Z, which is the metal burning luminosity, $\log$T, and $\log$P to be smaller than 5\%, and in $\log$L\_nuc and $\log$Rho, to be less than 10\% between timesteps. 
We reduce the timestep if the central abundances of C, O, or Ne change by more than 0.5\% between consecutive steps. We adopt \verb|gold2| solver tolerances and decrease \verb|varcontrol_target| ($w_t$), which is the target value for relative variation between consecutive timesteps, and \verb|time_delta_coeff|($\delta_t$) to $3\times 10^{-4}$ and 0.7, respectively. \\

\subsubsection{TP-AGB}

During the TP-AGB stage, we use the Schwarzschild criterion for convection. Overshooting is accounted for only at the bottom of the interpulse convective zone, with $f=0.008$ \citep{Herwig_2008}, because including convective-envelope overshooting in models that undergo hot third dredge-up (HTDU) can lead to numerical instabilities \citep{Rees+Karakas}.
We follow the recommendations of \citet{Rees+Karakas} for resolving the HTDU, adopting \verb|varcontrol_target| ($w_t$)= $2\times 10^{-5}$, \verb|mesh_dlogX_dlogP_extra|=0.8 for $^{4}$He, and \verb|delta_lgL_He_limit|=0.01. We increase the overall number of zones by setting \verb|mesh_delta_coef|=0.85, which roughly increases the number of zones by 15\%, and further increase resolution in He-burning and metal-burning zones by $\sim 40\%$. We switch to the \verb|eps_grav| form of the energy equation, which is better suited for degenerate conditions as it reduces numerical errors to entropy \citep{mesa4}, and do not account for element diffusion. Hydrogen recombination instability (HRI) is handled by adopting the \citet{Rees+Izzard} subroutine.

We maintain the blocker formulation for mass loss with the same $\eta_B$, restricting $\dot{m}$ to a maximum of $10^{-3}\,\mathrm M_\odot/$yr, as this should be the maximum stable mass loss for AGB stars \citep{vanLoon1999,Hofner18_mdot}. % (maximum stable mass loss for RG or AGB star according to vanLoon+1999b) (Hofner+Olofsson Review AGB Mass-loss2018 -> more extreme objects have been found with mass losses over 10-4) 

\subsubsection{Exiting the TP-AGB stage and post-AGB evolution}

To bypass the instabilities that plague SAGB computations at the end of the TP-AGB phase (mainly FeI) and allow calculations to continue, we force an exit from the AGB after over 100 pulses have been calculated. To do that, we increase $\eta_B$ to 10, limiting $\dot{m}_{max}$ to $1\rm M_\odot/$yr, and, in addition, we adopt the MESA MLT++ treatment for superadiabaticity \citep{mesa2}, by setting \verb|okay_to_reduce_gradT_excess| = true. This produces a boost to the energy transport efficiency by reducing the temperature gradient (gradT) so that it is closer to the adiabatic in regions that are convective but radiation-dominated.

To avoid fully removing the H envelope, we gradually reduce $\dot{m}_{max}$ to $10^{-5}\,\rm M_\odot/$yr when the hydrogen envelope mass ($M_\mathrm{Henv}$) falls below 0.1\,$\rm M_\odot$, completely stopping it once $M_\mathrm{Henv} < 10^{-5} \rm M_\odot$. After this point, any reduction in the H mass is due to burning.      

We revert to the Ledoux criterion during the post-AGB phase and consider thermohaline convection with $\alpha_{\rm th}=3$, as we find that this choice helps to smooth element discontinuities. We account for element diffusion and employ the formulation by \citet{Caplan+2022} for strongly coupled regimes.

\subsection{Evolution on the WD cooling track}\label{subsec:wd_stage}
We calculate the evolution on the WD cooling track for models with both hydrogen and helium-dominated atmospheres. 
To generate white dwarf models with hydrogen-dominated atmospheres, we turn off nuclear reactions when $M_{\mathrm H} < 1.45 \times 10^{-7}\, \rm M_\odot$. This value is extrapolated from Table~1 of \citet{lauffer}, and this approach prevents all of the hydrogen from burning, allowing us to bypass the need to add it at the start of the cooling curve.    

During WD evolution, we allow a maximum timestep of 1\,Myr. Convection is calculated with the ML2 \citep{ML2} formalism for MLT, with $\alpha_{\rm MLT}=0.8$ \citep{Gianninas11}. The Ledoux criterion determines the convective regions prior to crystallization, whereas the Schwarzschild criterion applies from the onset of crystallization onward. We further extend the convective boundaries by adopting exponential overshooting at the bottom of the convective shell with f=0.01, and thermohaline with $\alpha_{\rm th}=10$.  

Atmospheric boundaries above 40\,000~K are calculated using the gray Eddington t\,(tau) relation with variable opacity. Below 40\,000 K, we adopt the atmosphere tables for DA and DB white dwarfs provided by MESA \citep{Rohrmann11,Koester20}.
We account for element diffusion by adopting the formulation of \citet{Caplan+2022} for strongly coupled regimes and also consider the effects of element sedimentation.

Crystallization is treated self-consistently by Skye \citep{Skye}, using free energy minimization. In order to  handle phase transitions, Skye determines a smooth phase parameter ($\phi$\footnote{$\phi = \frac{e^{\Delta f/w}}{e^{\Delta f/w} + 1}$, where $f$ is the free energy of an ion and $w = 10^{-2}$ is a blurring parameter}), where $\phi = 0$ corresponds to a fully liquid regime, $\phi =1$ to fully solid, and $\phi=0.5$ marks the onset of crystallization
. Phase separation, initially implemented in MESA by \citet{Bauer2023_PS}, is calculated using \citet{Blouin2021_ONe_PS} phase diagrams for O/Ne mixtures. The solid mixing free energy formula we adopt in Skye is that of \citet{Ogata93}, and we account for latent heat release and phase-separation heating. We choose to extend plasma fits by setting \verb|skye_min_gamma_for_solid|=100 and \verb|skye_max_gamma_for_liq|=300 (see \citet{Skye} Figure 8), as those are values previously adopted in MESA when handling phase separation \citep{Bauer2023_PS} and ultramassive WD cooling \citep{Schwab_2021}.

\citet{Schwab_2021} calculated cooling models for O/Ne white dwarfs in the mass range 1.29 to 1.36 $ \rm \rm M_\odot$ with a fixed composition. He found that Urca cooling becomes active at masses $\gtrsim 1.33 \rm \rm M_\odot$. \citet{Althaus2022} calculated O/Ne WD models with a fixed composition, modifying the equations of stellar structure and evolution to include the effects of general relativity (GR). At $1.31 \rm \rm M_\odot$, they find that the inclusion of GR decreases the WD radius and cooling time by $\sim\,$5\%. Therefore, we do not include in our computations the effects of general relativity or neutrino loss via the Urca process, as these contributions are expected to be negligible at our final mass. 

\section{Results}\label{sec:results}
We evolve a 9\,$ \rm \rm M_\odot$, Z=0.02 model starting from the pre-ZAMS until it cools to $\sim10\,000$~K on the white dwarf cooling track. It results in a $\sim 1.313\,\rm M_\odot$ WD, with an ONe core.

We summarize the main aspects of our sequence throughout its evolution in Table~\ref {tab:geral}. We adopt the default definitions of H-free core ($\mathrm{M_{He}}$) and He-free core ($\mathrm{\rm M_{CO}}$) from MESA, which correspond to the mass of the outermost location where the abundances, by mass, of $^{1}$H and $^{4}$He, respectively, are less than 10\%.   

\begin{table}
    \centering
    \begin{tabular}{|lr|} \hline
    
     $\rm M_{He}^{CHeB}$ &$2.328\, \rm M_\odot$ \\\hline
     $\rm M_{CO}^{CHeB}$ & $1.109\,\rm M_\odot$\\ \hline
     $\rm M_{CO}^{\rm C\,ignition}$ &  $1.258\,\rm M_\odot$\\ \hline
     $\rm M_{CO}^{\rm 1^{st} thermal \,pulse}$ & $1.310\,\rm M_\odot$\\ \hline
     WD Final Mass& $1.313\, \rm M_\odot$  \\ \hline
     Age$^{\rm CHeB}$& $31.2\times10^6$yr\\ 
     \hline
     $\rm t_{cool}$ (10\,000~K)&   $1.667\times10^9$yr\\ 
     \hline
     Total Age (10\,000~K)& $1.698 \times10^9$yr\\ 
     \hline
     
    \end{tabular}
    \caption{Relevant aspects of the baseline model at different evolutionary stages.$\rm M_{He}^{CHeB}$, $\rm M_{CO}^{CHeB}$, and Age$^{\rm CHeB}$ are the H-free and He-free core masses, and the total age at the end of the core helium burning stage. $\mathrm{M_{CO}^{C \, ignition}}$ and $ \rm M_{CO}^{\rm 1^{st} thermal \,pulse}$ are the mass of the He-free core at the time of carbon ignition and at the start of the first thermal pulse, respectively. WD Final Mass is the final stellar mass when the model enters the WD cooling track. $\rm t_{cool}$ (10\,000~K) and Total Age (10\,000~K) are, respectively, the WD cooling time and total stellar age when the WD reaches $\rm T_{eff} =10\,000$K}
    \label{tab:geral}
\end{table}

\begin{figure}
    \centering
    \includegraphics[width=\linewidth]{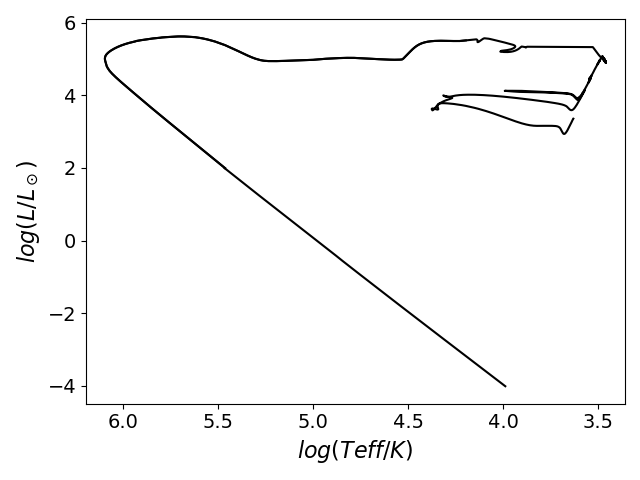}
    \caption{Hertzsprung-Russel (HR) Diagram following the evolution of a sequence with $\rm M_\mathrm{ZAMS} = 9\, \rm M_\odot$ and Z=0.02, from the zero age main sequence (ZAMS) to the WD cooling track }
    \label{fig:HRD}
\end{figure}

Figure~\ref{fig:HRD} presents the HR diagram for our sequence, including the pre-ZAMS and the WD cooling curve. We note that the plateau between $\log \rm T_\mathrm{eff}=3.5$ and 4.0, at $\log$L $\sim 5$, is the result of artificially removing a luminosity bump caused by our approach to leaving the TP-SAGB stage. 
\begin{figure}
    \centering
    \includegraphics[width=\linewidth]{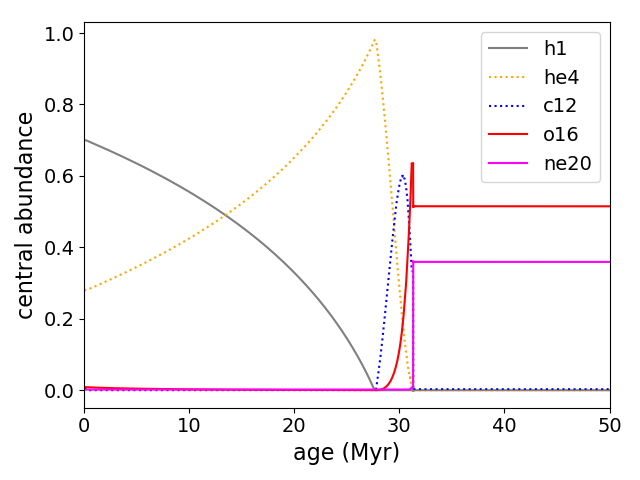}
    \includegraphics[width=\linewidth]{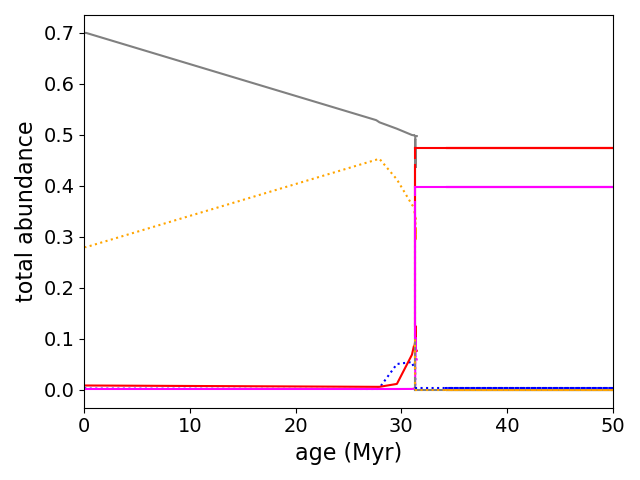}
    \caption{Evolution of central (top panel) and total (bottom panel) abundances of $^1$H (grey), $^4$He (dotted orange), $^{12}$C (dotted blue), $^{16}$O(red), and $^{20}$Ne (pink) from MS to the white dwarf stage, before phase separation sets in. Once phase separation occurs, the central fractions of oxygen and neon change.}
    \label{fig:chem_ev}
\end{figure}

In Figure~\ref{fig:chem_ev}, we show the evolution of the central and total abundances , both in mass fractions, of our main chemical species. To make the evolutionary phases clear, we present only the first $20$\,Myr of the white dwarf stage. Once phase separation occurs, the central fractions of $^{16}$O and $^{20}$Ne change, but $^{16}$O remains the overall dominant species.

Over the past 30 years, various studies have calculated sequences with the same mass and metallicity as those employed here. It is therefore of interest to briefly examine how our results compare to those of previous works. \citet{Garcia-Berro_R_I_9Msol} terminated their calculations after 7 thermal pulses and found $\rm M_{\rm CO} = 1.12\rm M_\odot$, with an ONe core. Their model leaves the CHeB stage at a younger age than ours, with only $\sim 23$\,Myr. The 9\,$\rm M_\odot$ models from both \citet{Siess2006} and \citet{Doherty2010} are similar to that from \citet{Garcia-Berro_R_I_9Msol}, which is expected, as they did not include convective boundary mixing during CHB and CHeB. At the time they terminate their calculations, at the first TP, both works have an ONe core and $\rm M_{\rm CO}$ = 1.063\,$\rm M_\odot$ and 1.099\,$\rm M_\odot$, respectively. Other works either find results similar to those or find that a sequence with initial mass and metallicity of 9\,$\rm M_\odot$ and $Z=0.02$ no longer produces a WD, terminating its evolution as a supernova \citep{Poelarends_2008,Woosley_Heger2015}. The exceptions are the works of \citet{lauffer}, for which they find a CO WD with 1.024\,$\rm M_\odot$, \citet{doherty+2015}, which finds $\rm M_{\rm CO}$ = 1.225\,$\rm M_\odot$ and an ONe core, at the end of the TP-SAGB, and the K9 sequence from \citet{Poelarends_2008}, which has $\rm M_{\rm He}=1.338\,\rm M_\odot$ after the second dredge-up (2DU) and based on their synthetic TP-AGB evolution is expected to form an ONe WD. The sequence from \cite{doherty+2015} is very similar to the computations in this work, as it starts with the same mass and metallicity, and has an age of $\sim 29$\,Myr at the end of CHeB. However, it must undergo carbon burning under fairly different conditions, given that, at ignition, the sequence computed in this work already has a larger $\rm M_{\rm CO}$ than their final model.

\subsection{Carbon burning/EAGB}
At the beginning of the EAGB, $\rm M_{\rm He}$ and $\rm M_{\rm CO}$ are 2.328\,$\rm M_\odot$ and 1.109\,$\rm M_\odot$, respectively, and the total stellar mass is 8.91\,$\rm M_\odot$. At this point, the central abundances in mass fractions are $X_{\rm C}$=0.33 and $X_{\rm O}$=0.63. It is only after $\sim 0.12$\,Myr, since the end of CHeB, that Carbon ignites in a flash, %only after $\sim 0.12$\,Myr of He-shell burning,
when $ \rm M_{\rm CO} = 1.259\, \rm M_\odot$. 

\par
\begin{figure}
    \centering
    \includegraphics[width=\linewidth]{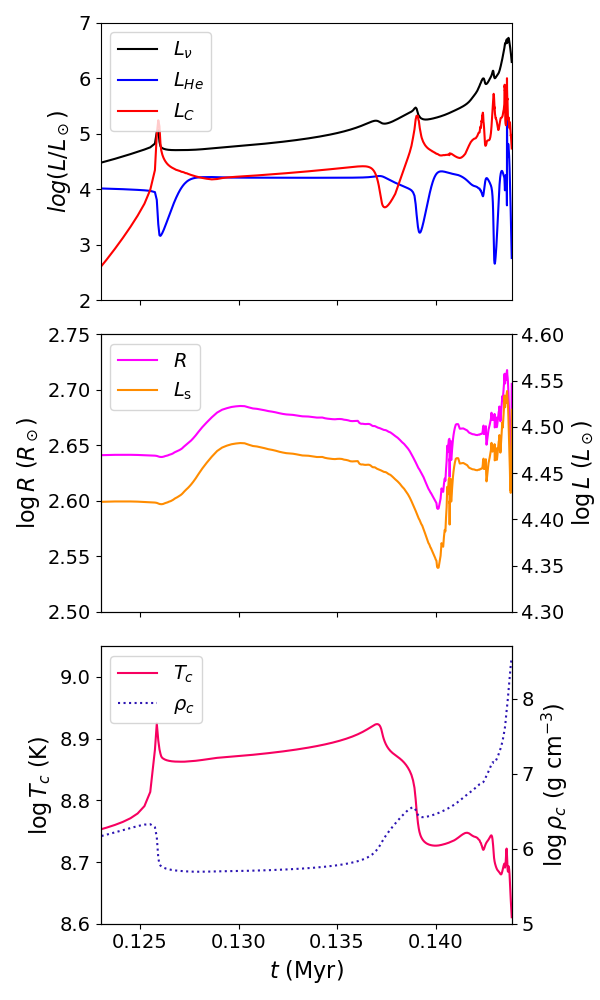}
    \caption{Top Panel: Helium luminosity $\rm L_\mathrm{He}$ (black), carbon luminosity $\rm L_\mathrm{C}$ (red), and neutrino luminosity $\rm L_\nu;$ (blue) during C-burning. Middle Panel:  Radius (pink) and Surface Luminosity (orange). Bottom panel: central temperature (pink) and density (blue dotted). The time axis is reset so that zero marks the start of the EAGB.}
    \label{fig:cburn}
\end{figure}

\begin{figure}
    \centering
    \includegraphics[width=\linewidth]{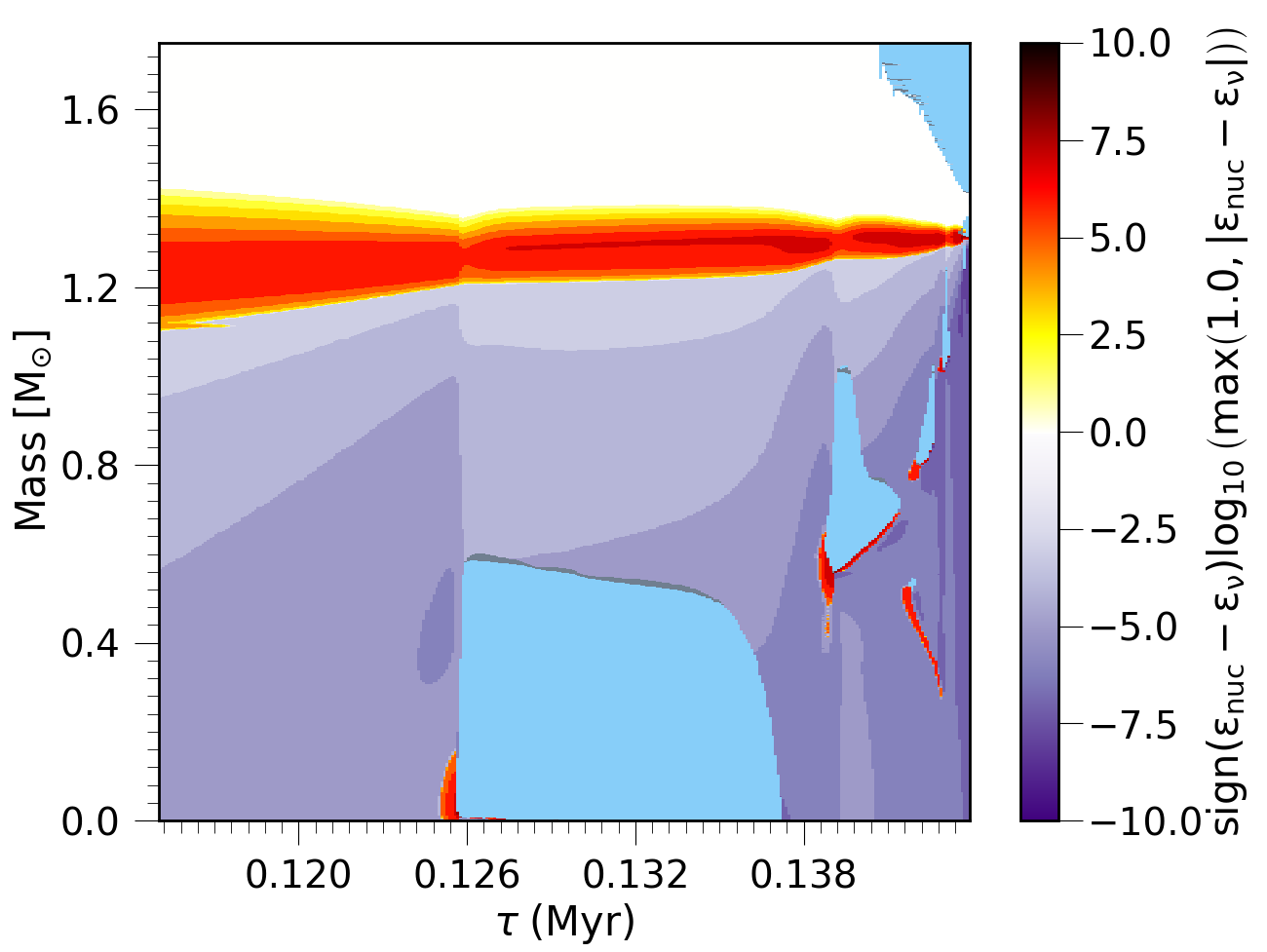}
    \caption{Kippenhahn Diagram following carbon burning in the EAGB. The color bar represents the same quantity used in the Kippenhahn plots of \citet{timmes+2015}, where purple regions indicate cooling, primarily due to thermal neutrino losses. Darker purple indicates a logarithmic increase in thermal neutrino production rates.
        Yellow/orange regions indicate significant nuclear burning, with redder colors indicating a logarithmic increase to the nuclear reaction rates; light blue regions indicate convection. For clarity, in this plot we show only the inner 1.75\,$\rm M_\odot$, and the time axis is reset so that zero marks the start of the EAGB.}
    \label{fig:KipD}
\end{figure}

 The energy produced by Carbon and Helium burning, as well as the energy lost by thermal neutrinos, can be seen in the top panel of Figure~\ref{fig:cburn}, which covers the entire carbon-burning stage. In the same figure, we also present local characteristics of the surface (middle panel) and core (bottom panel). We note that from the beginning of the first flash to a little before the start of the second, $\rm T_c$ assumes the same value as $ \rm T_{\rm max}$. After this point, $ \rm T_{\rm max}$ moves outward, tracking the location of the subsequent flashes, which occur progressively closer to the He shell. 
We can see both the time and location of the following flashes in the form of a Kippenhan diagram in Figure~\ref{fig:KipD}. Convective zones are represented in light blue, grey indicates semiconvection, and red represents significant nuclear burning. In dark purple are the regions of intense thermal neutrino losses. 


\par As can be seen in Figure~\ref{fig:KipD}, only after the second flash does the efficiency of He-burning significantly decrease, reducing the $ \rm M_\mathrm{CO}$ growth and allowing the convective envelope to start moving inward. However, before it can fully travel and reach its innermost zone, a semiconvective and later fully convective region is established in the He-burning shell, causing a dredge-out to occur. Similarly to what was found by \citet{doherty+2015}, the He-shell is extinguished and the convective envelope penetrates the outward regions of the CO core, slightly reducing $M_\mathrm{CO}$ by $\sim 0.002\, \rm M_\odot$. Additionally, for a brief period, no region is dominated by Helium.

To verify that our baseline model accurately captures the carbon-burning stage, we conducted additional computations to assess the impact of increasing temporal and spatial resolution. In those tests, we only changed \verb|mesh_delta_coeff| ($\delta_m$), \verb|varcontrol_target|($w_t$), and \verb|max_years_for_timestep|. We present in Table~\ref{tab:eagb_res} the initial and final global quantities relevant to the C-burning stage. The final properties are taken before the occurrence of the dredge-out, when $\rm M_\mathrm{He} - \rm M_\mathrm{CO}\leq 0.015\,$$\mathrm{M_\odot}$, while the initial properties are taken during the first C-flash, at the time of the highest Carbon luminosity ($\rm L_C$). 
Test1 has $\delta_m$=0.75 which roughly corresponds to a 25\% increase in the number of zones, Test2 has $w_t=10^{-4}$, and Test3 has $\delta_m =0.5$ , which lends to about twice as many zones as the baseline model, and $w_t=10^{-4}$, which are the same values used by \citet{timmes+2015}. In addition, we impose a maximum allowed timestep of 100\,yr for tests and 2. As a result of the increased resolution, Test1  and Test3 have an average of respectively  $\sim $ 700 and $\sim$1700 more zones in the core region than our baseline model. Additionally, they have $\sim $400 and $\sim $650 more zones in the first c-burning region. As for the increase in temporal resolution, Test2 and Test3 require  about 1000 more timesteps  to follow the entire carbon-burning stage.

\begin{table}
    %\centering
     \raggedleft\begin{tabular}{|c|c|c|c|c|} \hline
     run &  $\rm M_{CO}^{ign}(\rm M_\odot) $ & $\rm M_{CO}^{f}(\rm M_\odot)$& $Age^{f}(yrs) $& $X^f(\mathrm{C^{12}}_{center})$\\ \hline
      baseline    & 1.258& 1.312 & $1.438\times10^5$ &$0.0028$\\
      Test1 &1.258&1.309& $1.434\times10^5$&$0.0020$ \\
      Test2 &  1.259&1.310&$1.440\times10^5$&$0.0029$ \\ 
      Test3 & 1.259 & 1.310&$1.436\times10^5$& $0.0019$\\ \hline

    \end{tabular}
    \caption{Initial ($\rm ^{ign}$) and final ($^{f}$) global properties for different spatial and temporal resolutions during C-burning. Test1 has $\delta_m$=0.75, Test2 has $w_t=10^{-4}$, and Test3 has $\delta_m =0.5$ and $w_t=10^{-4}$. In addition, we impose a maximum allowed timestep of 100\,yr for tests 1 and 2.}
    \label{tab:eagb_res}
\end{table}

The choice of presenting the initial quantities in Table~\ref{tab:eagb_res} at the time of largest $L_C$ is motivated by the fact that it is straightforward to define. The precise time and location of Carbon ignition are less clear, as they depend on the adopted criteria and model resolution. Furthermore, historically, different works have employed varying or imprecise definitions (e.g., ``strong/vigorous burning'' without a numerical quantity attached) for C-ignition, making model comparison and criterion selection cumbersome. This is especially true for core masses that sit at the edge of the center/off-center ignition transition, such as the one computed in this work. We illustrate this in Table~\ref{tab:eagb_ign_comparison}, where we present local quantities at ignition for different $\epsilon$/$\nu$ ratios. The location is always the innermost cell where $\mathrm{X(^{20}Ne)>X(^{23}Na)>X(^{24}Mg})$ and the condition of column 1 is satisfied.  

\begingroup
\renewcommand{\arraystretch}{1.1}
\begin{table*}
    \centering
    \begin{tabular}{l|rrr|rrr} 
    \toprule
     & \multicolumn{3}{c}{baseline}&\multicolumn{3}{c}{test3} \\
     \cmidrule(lr){2-4}\cmidrule(lr){5-7}
     %\midrule
      $\epsilon/\nu \geq$& $m_{ign}(\rm M_\odot)$ & $log(\rho_{ign})$ & $T_{ign}(10^8K)$ & $m_{ign}(\rm M_\odot)$ & $log(\rho_{ign})$ & $T_{ign}(10^8K)$ \\
     \midrule
     1 & 7.0d-7&6.32&6.5&3.5d-7&6.32&6.1\\  %\hline
     2 & 7.0d-7&6.28&7.6 & 3.5d-7&6.32&6.5\\  
     3 & 8.5d-3&6.26&7.4 &3.5d-7&6.27&7.0\\  
     4 & 7.0d-7&6.15&8.4& 3.4d-7&6.27&7.0\\  
     5 & 7.0d-7&6.15&8.4 & 8.4d-3&6.22&7.6\\  
     9 &7.0d-7&6.15&8.4 & 1.0d-3&6.10&8.4\\  
     \bottomrule
    \end{tabular}
    \caption{Ignition properties as a function of different required $\epsilon/ \nu$ ratios. The ignition location is always the innermost cell where $X(^{20}\rm Ne)>X(^{23}\rm Na)>X(^{24}\rm Mg)$ and the condition of column 1 is satisfied.} 
    \label{tab:eagb_ign_comparison}
\end{table*}
\endgroup

\par
\begin{figure}
    \centering
    \includegraphics[width=\linewidth]{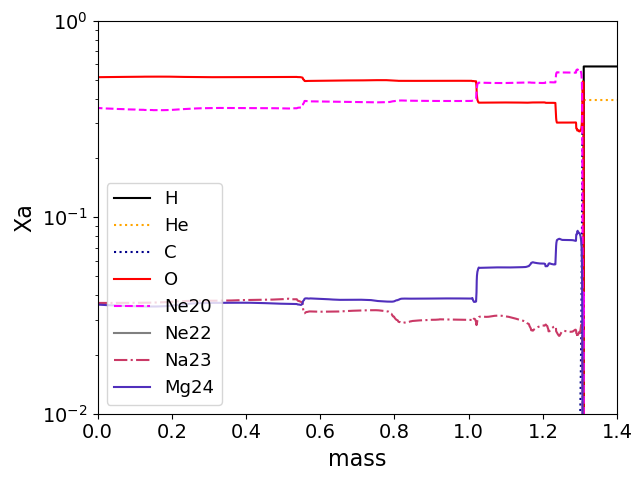}
    \caption{Abundance profile in the inner 1.4$\,\rm M_\odot$ for our main species right after the dredge-out has occurred.}
    \label{fig:eagb_profile}
\end{figure}

 At the end of carbon burning, $^{16}$O and $^{20}$Ne are the most abundant species in the core, corresponding to 47.4\% and 39.7\% of the total core mass, respectively. This difference is larger in the center, with ${X_{^{16}\rm O}=0.514} $, it decreases at $m_r\sim 0.55 \rm M_\odot$, and finally switches at $m_r \sim 1.0\,\rm M_\odot$, with $^{20}$Ne becoming the dominant species in the outer $\sim 0.3\rm M_\odot$ of the core. Similarly, $^{23}$Na is slightly more abundant, by mass fraction, than $^{24}$Mg in the innermost 0.55$\,\rm M_\odot$; this switches at this point, and at $m_r \sim 1.0\,\rm M_\odot$, $^{24}$Mg becomes significantly more abundant than $^{23}$Na, making it our third most abundant species. This does not appear to be a resolution-related effect, as the behavior is observed across all tests. We show in Figure \ref{fig:eagb_profile} an abundance profile after the occurrence of the dredge-out. A similar abundance distribution is also seen in Figure~32 of \citet{Iben_Ritossa_GarciaBerro_4}, for a 1.263$\,\rm M_\odot$ core model. However, in that figure, the $^{16}$O and $^{20}$Ne profiles only approach each other and do not cross.    

We proceed with the evolution of our baseline model, as we find little difference in the global properties across our tests. The largest discrepancy is in the final central abundance of $\mathrm{^{12}C}$, for which our baseline model finds $\rm X_C = 0.00280$, while Test3, which has the highest resolution among our tests, has $\rm X_C = 0.00195$. This discrepancy is likely due to the additional spatial resolution, since Test1 presents a similar $\rm X_C$ to Test3.

Although there is still no consensus in the literature on the impact of rotation on degenerate carbon burning, it is unlikely that at this stellar mass the inclusion of rotation during the EAGB could have prevented carbon from burning. Works such as \citet{Althaus2021} find that rotation prevents all but the most massive SAGB stars from undergoing degenerate carbon burning. While others report it to have no impact on the threshold mass for carbon ignition \cite[e.g.][]{timmes+2015}. However, there is still a threshold mass above which carbon will ignite. Our model surpasses in mass both the classic limit $\rm M_{CO} > 1.05 \rm M_\odot$ in the EAGB and the one set by \citet{Althaus2021}, which seems to be $\rm M_{He}> 1.735 \rm M_\odot$, at the end of CHeB.

The uncertainties in the $^{12}\rm{C}+^{12}\rm{C}$ reaction rate and its channels are also unlikely to prevent carbon burning in our model. \citet{deGeronimo2023} investigated this precise issue, finding that it could only produce UMWD with CO cores up to a mass of $\sim$1.10$\rm \rm M_\odot$. Although there is observational evidence for CO UMWD, we find it unlikely that those stars descend from such a high-mass progenitor with extended core convection, such as the one we explore in this work. It is possible that the extreme modification of the $^{12}\rm{C}+ ^{12}\rm{C}$ reaction rates coupled to a rotation scheme such as the one implemented by \citet{Althaus2021} could prevent a model like ours to undergo carbon burning, however, the only sure way for our model to result in a CO UMWD would be to artificially turn off carbon burning reactions. 

\subsection{TP-AGB} 

We calculate the evolution in the TP-SAGB for 141 pulses, which occur within $\sim 3000$~yr. During this time, $\rm M_\mathrm{CO}$ goes from 1.3104 to 1.3126\,$\rm M_\odot$, with a remaining envelope mass of 3.91\,$\rm M_\odot$.

\begin{figure}
    \centering
    \includegraphics[width=\linewidth]{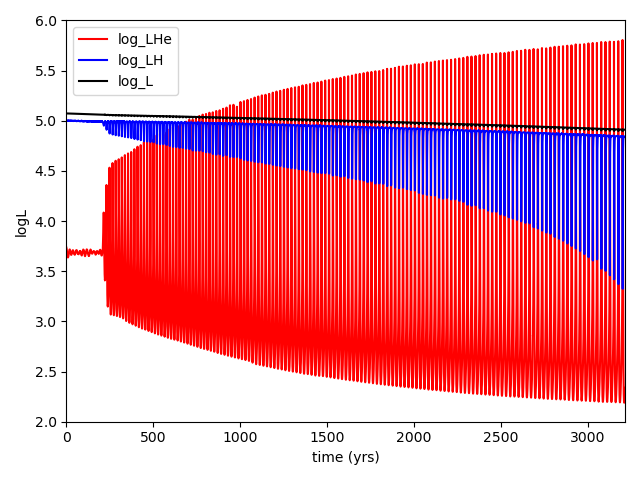}
    \caption{Surface (black), He (red), and H (blue) luminosities as a function of time in the TP-SAGB stage. In this plot, time is reset so that the first He-shell instability corresponds to a time of zero.}  
    \label{fig:tp_age}
\end{figure}

Figure~\ref{fig:tp_age} shows the He, H, and surface luminosities from the first He-shell instability ("mini-pulse") to the last calculated thermal pulse. It takes almost 1000~yr for the helium luminosity to exceed $\log$L, and although the pulses become progressively stronger, $\rm L_\mathrm{He}$ never reaches $10^{6.0} \,\rm \rm L_\odot$ during our calculations. This may be a result of truncating the TP-SAGB, but other works that carried out TP calculations also found this feature in their more massive models \citep{siess2010,Karakas_SNhand}, so this could be a consequence of their weaker pulses.

\begin{figure}
    \centering
    %\plotone{figures/tpagb_core_age_zoom2.png}
    \includegraphics[width=\linewidth]{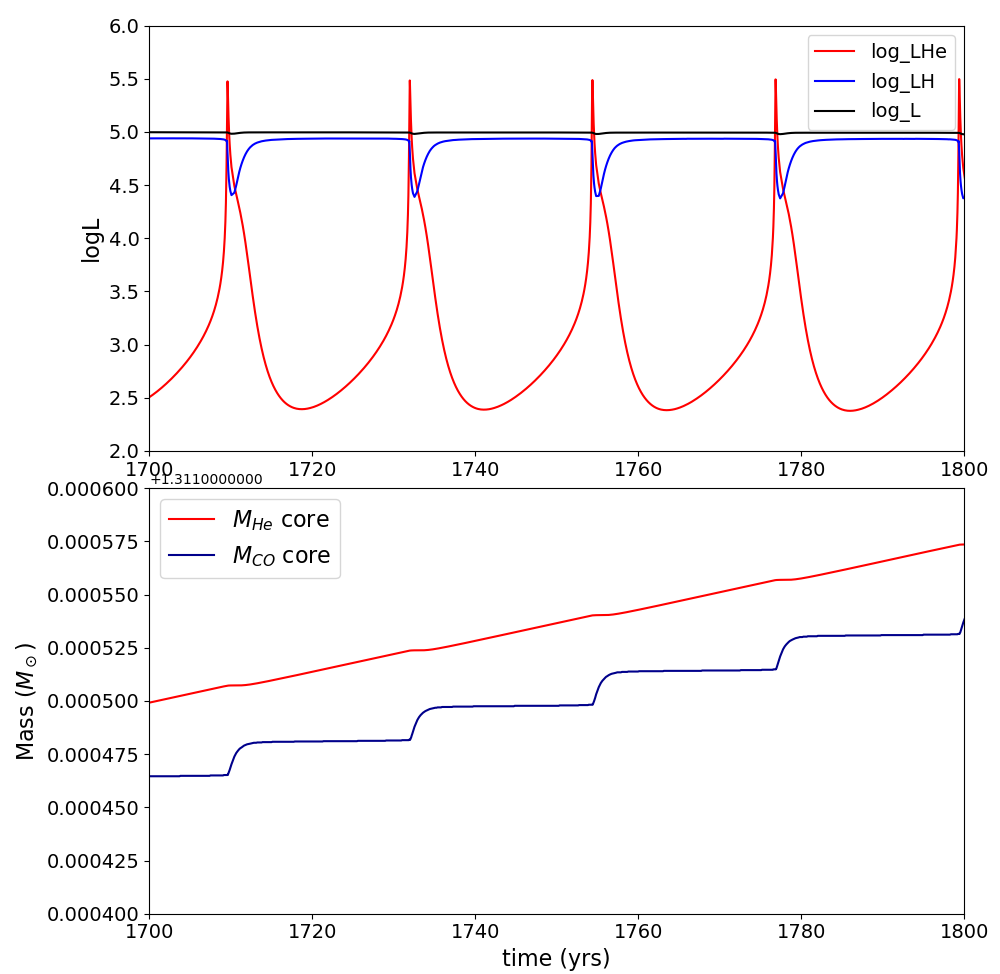}
    \caption{Top Panel: Surface (black), He (red), and (blue) H luminosities as a function of time between pulses 68 and 72.
    Bottom Panel: Masses of the H-free (red) and He-free (blue) cores during the same interval shown in the top panel.} 
    \label{fig:tp_zoom}
\end{figure}

Like in all massive SAGB stars, the interpulse period in our model is much shorter than in typical low-mass AGB stars, with an average duration of $\sim 20$~yr. For comparison, the typical interpulse period for low-mass AGB stars is between $10^{4}$-$10^{5}$~yr. The pulse-interpulse regimes can be better seen in Figure~\ref{fig:tp_zoom} for pulses 68 to 72. In the top panel, we show $\rm L_\mathrm{H}$, $ \rm L_\mathrm{He}$, and $\rm L_\mathrm{surf}$, with the same color scheme as Figure~\ref{fig:tp_age}, while in the bottom panel, we present the masses of the H and He-free cores at the time. There, the $\rm M_\mathrm{CO}$ growth during each thermal pulse is evident, if relatively small, but the effects of the third dredge-up are much more subtle, as we find it to be very inefficient, with the maximum 3DU efficiency\footnote{ $\lambda = \Delta \rm M_{dredge}/ \Delta M_{He}$, where $\Delta\rm M_{He}$ is the mass increase to the H-free core during the previous interpulse period and $\Delta \rm M_{dredge}$ is the mass dredged up from the intershell region.} $\lambda ^\mathrm{max} \sim 0.02$. Although this may be an effect of our high mass-loss rate \citep{Rees+Karakas}, 3DU efficiency varies wildly between codes, and there are no observational constraints for it in SAGB stars \citep{Poelarends_2008,Karakas_SNhand}. Given that different codes generally assume different physics that go beyond different wind prescriptions, a systematic exploration of mass-loss effects on 3DU is needed to confirm the suggestion that fast winds lead to small $\lambda$. 

We choose to force an exit of the TP-SAGB after the $139^\mathrm{th}$ pulse, since from this point onward, the instabilities associated with the end of the TP-SAGB progressively grow, making each consecutive thermal pulse more computationally expensive and less numerically stable. These instabilities have been extensively discussed in the literature, and we refer the reader to the works of \citet{Lau+2012}, \citet{doherty+2015}, and \citet{Rees+Izzard} for in-depth discussion.

We have experimented with other strategies to overcome the Fe-peak instability, but find the one described in Subsection~\ref{subsec:prewd} to be the best suited to our interests. Approaches like the one presented in \citet{Lau+2012} (gradually increasing $\alpha_{\rm MLT}$) do work, but require significantly more computational time, demand a lot of ``fine-tuning'', and also result in less stable models in the post-AGB, which makes the transition to the white dwarf track a herculean task. As for the subroutine presented by \citet{Rees+Izzard} (locally increasing $\alpha_{\rm MLT}$ in the cells where $P_{\rm gas}/P$ becomes small), while it is elegant and less disruptive to the upper layers, in high-mass SAGB models, it drives the timesteps to such a small scale, (minutes and later seconds), that it becomes impractical, even after adopting an extremely high mass-loss rate. In addition, we also experimented with reducing $\dot{m}_{max}$, adopting the limits of $0.5 \rm M_\odot/$yr and $0.1 \rm M_\odot/$yr, finding no significant changes, except for the computational time required to reach the post-AGB.

We reiterate that, to date, there are no reliable hands-off methods to solve or work around the problems introduced by the Fe-peak Instability. And, as a consequence, there are no computational works on SAGB stars that manage to calculate \textbf{all} of the thermal pulses these stars are expected to undergo. Generally, at termination, the model will still have over $1\rm M_\odot$ of the H envelope, with this number being larger for higher masses and metallicities.  

To verify that our approach does not change the nature of the remnant, giving origin to a neutron star instead of a WD \citep{Nomoto1984,Poelarends_2008}, 
%significantly impact the structure of the white dwarf remnant,
we performed simple estimations under the assumption that the average $\Delta \rm M_\mathrm{CO}$ remains constant. Given $\Delta \overline{\rm M}_\mathrm{CO} = 7.3\times 10^{-7}\,\rm \rm M_\odot/$yr, $\dot{\rm M} =10^{-3}\,\rm M_\odot/$yr, and the remaining mass of hydrogen envelope, $\rm M_H = 3.91\, \rm M_\odot$, we estimate a duration of another $\sim$3910\,yr in the TPAGB, and an additional core growth of only $\sim0.003\,\rm M_\odot$, for a total core mass of ~1.315\,$\rm M_\odot$. If we were to impose a more modest mass-loss rate ($10^{-4}\,\rm M_\odot/$yr), the extra growth would be around 0.028\,$\rm M_\odot$, putting our core at $\sim1.34\rm M_\odot$. % Restricting mass-loss to this maximum throughout the entire tp-agb stage would result in a final core mass of ~1.36$\rm M_\odot$, right at the edge of the limit for electron-capture supernovae (Nomoto87,Takahashi2013).be no
As such, adopting our solution to leaving the TP-SAGB prematurely does not alter the nature of the remnant and causes only a slight decrease in its final mass. Furthermore, it might be physically consistent, given that a mass-loss boost or even total envelope ejection cannot be excluded as a possible consequence of Fe\,I \citep{Wood+Faulkner86,Sweigart1999,Lau+2012}. 

Entirely bypassing the TP-SAGB stage by early adoption of high winds would shorten the core mass only by a maximum of $\sim0.05\,\rm M_\odot$, compared to calculating the entire stage with a constant mass-loss rate of $10^{-4}\,\rm M_\odot/$yr. This is similar to the results of \citet{doherty+2015}, which reports a maximum core growth during TP-(S)AGB of $\sim0.01 - 0.03$, given their much more efficient 3DUs. Therefore, entirely skipping TP-SAGB calculations imposes only a very minor impact on the IFMR of ultramassive white dwarfs.

To explore whether skipping the TP-SAGB stage might result in white dwarfs with inaccurate Helium content, which has been suggested in the literature, we ran an additional sequence in which we skip thermal pulses by adopting a high mass-loss scheme ($\eta_B=10$) from the end of 2DU onward, until we stop it at $\rm M_H \sim 0.5\, \rm M_\odot$. We compare the thicknesses of the He-dominated ($\rm M_\mathrm{He}$) and C-dominated ($\rm M_\mathrm{C}$) layers at this point, finding that it increases $\rm M_\mathrm{He}$ by only a factor of 1.7 and decreases $ \rm M_\mathrm{C} $ by a factor 0.6. 

\begin{figure}
    \centering
    \includegraphics[width=\linewidth]{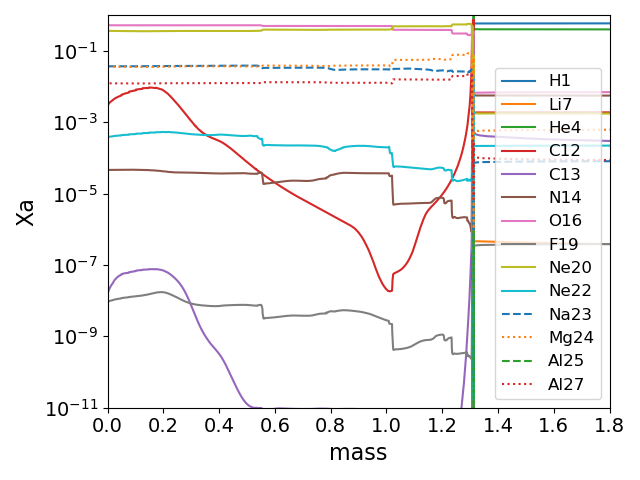}
    \includegraphics[width=\linewidth]{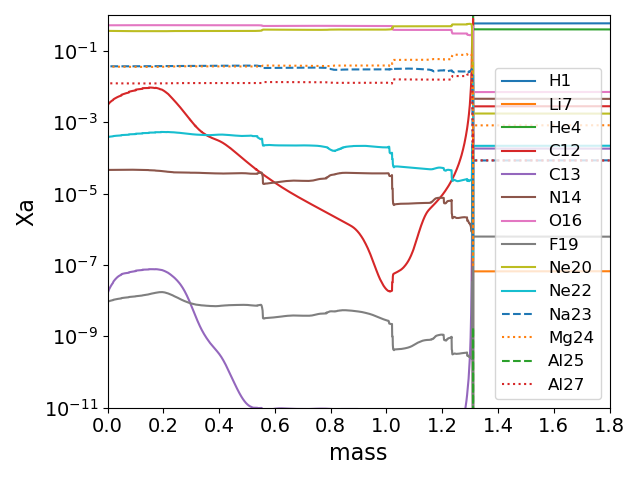}
    \caption{Abundance profiles when $\rm M_{\rm Henv} \sim 0.5$, for our baseline model with 139 thermal pulses (top panel) and for a model that bypasses thermal pulses due to the adoption of extreme mass loss (bottom panel)}
    \label{fig:TP_comp}
\end{figure}

Regarding the overall abundances changes, it can be seen in Figure~\ref{fig:TP_comp} that, except for the slightly larger $^{12}$C-dominated region in the model that undergoes thermal pulses, the composition differences are mainly in the envelope, the most noticeable being the abundances of $^{7}$Li, $^{12}$C, $^{13}$C, and $^{19}$F.

At the surface, the most significant changes are in the abundances of $^{7}$Li, $^{17}$O, $^{18}$O, $^{19}$F and $^{25}$Mg, which at $X_{\rm Li7} = 3.6\times 10^{-7}$, $X_{\rm O17} = 2.0\times 10^{-5}$,$X_{\rm O18} = 1.7\times 10^{-5}$, $X_{\rm F19} = 3.9 \times 10^{-7}$, and $X_{\rm Mg25} = 3.0\times 10^{-5}$ are 11 times higher in $^{7}$Li, $\sim$3 times higher in $^{17}$O and $^{25}$Mg, and about twice lower in  $^{18}$O and  $^{19}$F than for the sequence that does not undergo thermal pulses. There are also a few differences in surface fractions,  with the main being $^{12}\rm C/^{13}\rm C = 7.13$ vs 18.7 without TP, $^{12}\rm C/^{16}\rm O = 0.27$ vs 0.41, $^{17}\rm O/^{18}\rm O = 1.18$ vs 0.20, $^{24}\rm Mg/^{25}\rm Mg = 2.07$ vs 8.65 and $^{25}\rm Mg/^{26}\rm Mg = 2.14$ vs 0.68. 

Although these abundance differences in the envelope are significant for the study of (S)AGB stars and stellar populations \citep{Karakas2014_review}, as they affect the chemical yields, for the purpose of studying WD stars, they have minimal effects since the envelope will be almost completely stripped. Hence, we find that, at this mass, skipping TP-SAGB will result in only minimal alterations to both the final mass and the white dwarf chemical composition, making it a fast and reliable method for creating UMWD models. We caution that these results hold only if the adoption of high mass-loss rates happens \textbf{after} the 2DU, as we find significant core growth during the EAGB. 

\subsection{White Dwarf Evolution}\label{subsec:wd_results}
%\textcolor{green}{- Models say -> ultramassive = DB 
%-observations say: if it's ultramassive, it's DA. 
%gaia massive DBs -> actually massive DAH}
Despite stopping mass loss when the hydrogen envelope drops below $10^{-5}\, \rm M_\odot$, our baseline model leaves the post-AGB with virtually no remaining hydrogen ($\mathrm{M_H < 10^{-15} \,M_\odot}$), as all of it burns down. Since most %of the near-Mch
WDs observed are DAs \citep[e.g.][]{Obrien40pc},
we tried two different approaches to construct a hydrogen-rich WD: by adding H at the start of the cooling curve and by stopping nuclear reactions when $\rm M_H \sim10^{-7}\, \rm M_\odot$. Although we can cool these H-rich models past crystallization, we cannot account for diffusion or thermohaline processes, resulting in jagged abundance profiles and an envelope that is H-rich but He-dominated. We therefore focus our discussion in this section on our H-deficient sequences.

Our calculations result in a white dwarf with a mass of 1.3127\,$\rm M_\odot$, a total He mass of $1.5 \times 10^{-5}\rm M_\odot$ and an O/Ne/Mg core. The chemical abundance distribution in our model is such that 47.7\% of the total mass is $^{16}$O, 39.7\% $^{20}$Ne, 4.2\% $^{24}$Mg, 3.3\% $^{23}$Na, and 0.386\% $^{12}$C. The remaining $\sim 5\%$ are made up mainly of $^{25,26}$Mg and $^{27}$Al, with only 0.027\% being $^{22}$Ne.

\begin{figure}%[h]
    \centering
    \includegraphics[width=\linewidth]{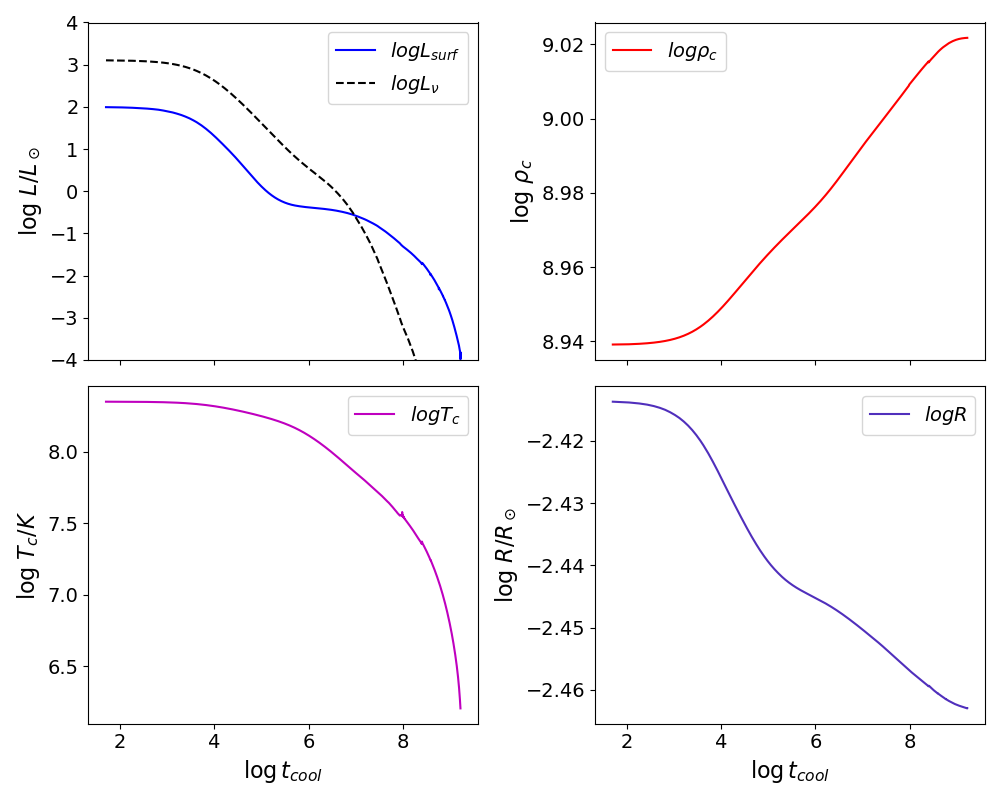}
    \caption{Physical quantities as a function of time, during the white dwarf cooling track. The top left panel displays surface (blue line) and neutrino (black dashed line) luminosities, while the top right panel shows the evolution of central density. The bottom left and right panels show central temperature and radius, respectively, alongside the cooling age.}
    \label{fig:coolWD4}
\end{figure}

We define the beginning of the white dwarf cooling track at $\rm L = 10^2 \, \rm L_\odot$, a little below the WD 'knee' in the HR Diagram, which for our sequences corresponds to an effective temperature slightly above 290\,000~K. Figure~\ref{fig:coolWD4} shows how the luminosities, radius, central temperature, and central density change as our model cools.% We tabulate a few of these quantities in Appendix~\ref{append:B}. 
\par
\begin{figure}
    \centering
    \includegraphics[width=0.90\linewidth]{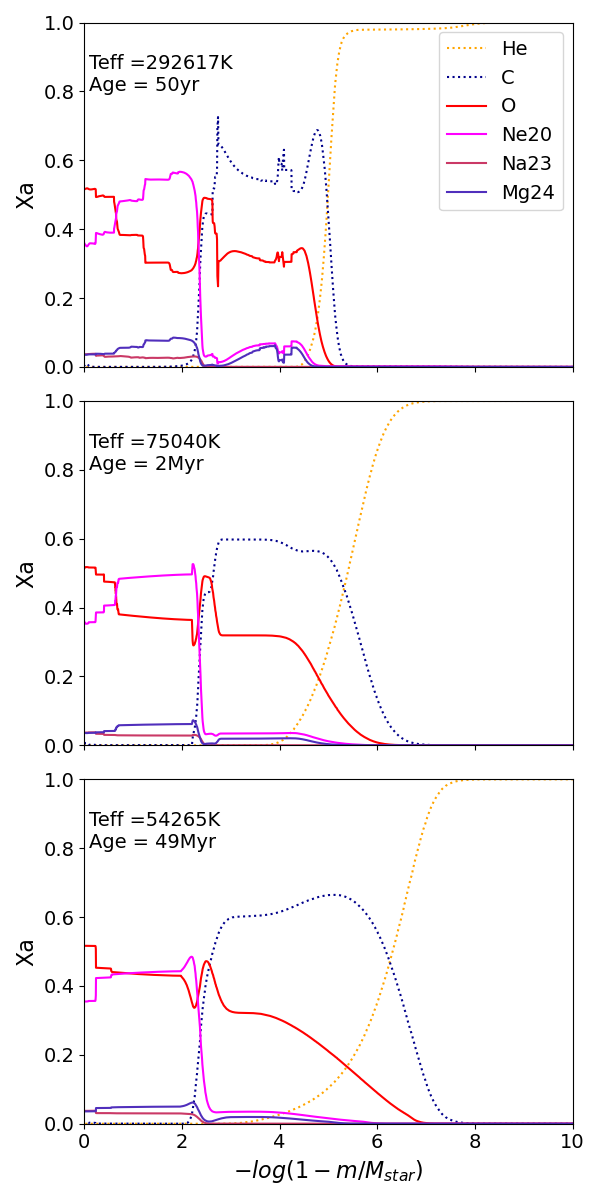}
    
    \caption{Abundance profiles for our white dwarf sequence at three different moments before the onset of crystallization. For clarity, we include in the plots only the most relevant species. Panel one shows the abundance profile at the start of the cooling curve. Panels ~2 and 3 show the profiles at $\mathrm{T_{eff}}= 75\,000$\,K and 54\,000K}
    \label{fig:pre_cryst}
\end{figure}

 As can be seen in the top right panel of  Figure~\ref{fig:coolWD4}, the first $\sim 10\,$Myr are dominated by neutrino cooling, during which the white dwarf luminosity drops to $\sim 10 ^{-0.6} \rm L_\odot$ and $\rm T_\mathrm{eff}$ to 69\,700~K.  
During this period of rapid cooling, element diffusion and thermohaline mixing are the main processes reshaping species distributions by smoothing the abundance profile, sinking heavier elements toward the center, and dragging lighter elements upward. Another $\sim 70$\,Myr pass before the onset of crystallization at 48\,530\,K, and by the time we stop our calculations, at L $\sim 10^{-4}\rm L_\odot$, our model has an effective temperature of 9772~K and a cooling age of 1.667\,Gyr. In Figure~\ref{fig:pre_cryst} we show how the abundance profile evolves before the onset of crystallization.

\begin{figure}
    \centering
    \includegraphics[width=0.90\linewidth]{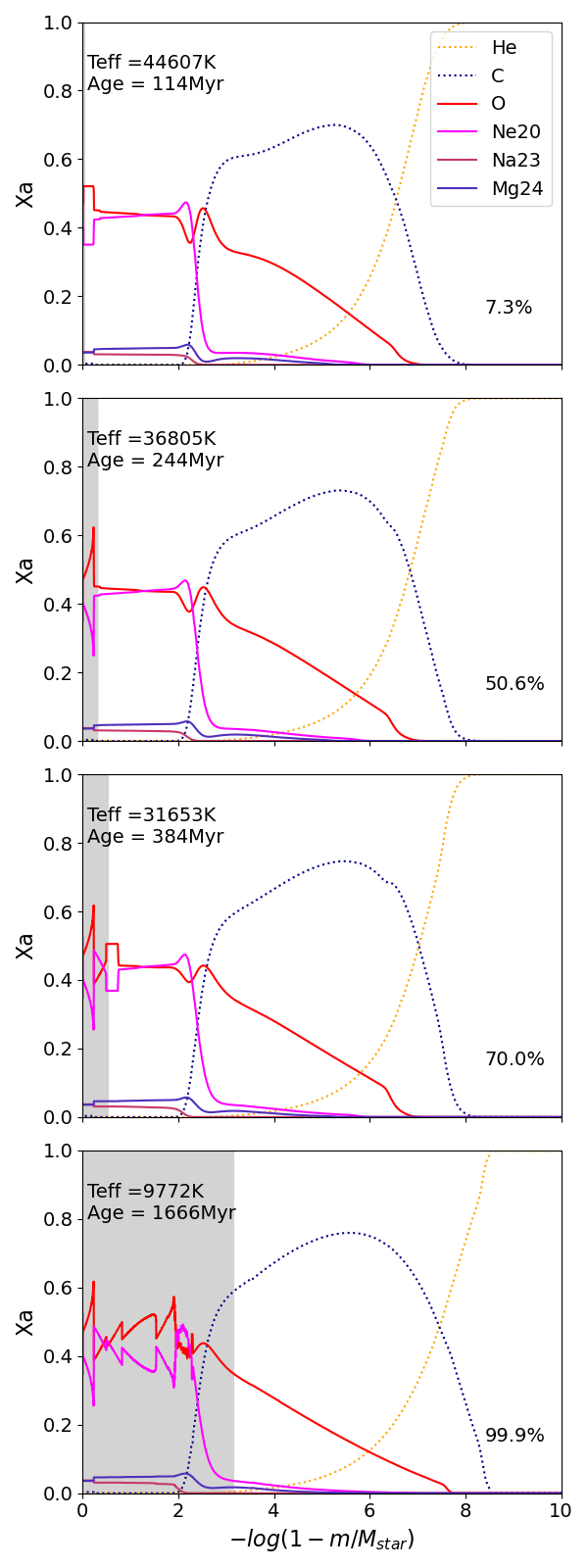}
    \caption{Abundance profiles for our white dwarf sequence with $1.313\,\rm M_\odot$ at four different moments in time. For clarity, we include in the plots only the most relevant species. Panel one shows the abundance profile near the onset of crystallization. Panels~2 and 3 show the profiles when approximately 50\% and 70\% of the stellar mass is crystallized (gray shaded area), at  $\rm T_\mathrm{eff}=36\,800$~K and 31\,650~K, respectively. We show in the final panel the profile of our last calculated model, at $\rm T_\mathrm{eff}=9772$~K.}
    \label{fig:wd_profile}
\end{figure}

It takes $\sim160$\,Myr from the beginning of crystallization to the point where 50\% of the star is crystallized, $\sim140\,$Myr from 50\% to 70\%, and $\sim270$\,Myr from 70\% to 90\%. Phase separation acts along crystallization, modifying the $^{16}$O and $^{20}$Ne profiles. Figure~\ref{fig:wd_profile} shows the abundance profiles at four different stages: slightly after the onset of crystallization, when the crystallization front reaches 50\% and 70\% of the star (in mass), and for our final model. At the time we stop our calculations, at $\rm T_\mathrm{eff}=9772$~K, only the upper $9.4\times 10^{-4}\,\rm M_\odot$ remains uncrystallized. The "zig-zag" behavior presented by the O and Ne profiles, which is more evident in the bottom panel, is not caused by crystallization alone (see left bottom panel of Figure~\ref{fig:wd_profile_NoPS} in \ref{append:B}). It, however, seems to be related to the MESA phase separation problem described in \citet{Castro-TapiaPS}, where phase separation heating causes the crystal to remelt, thus allowing for further mixing.

Recently \citet{Desai2025} published new determinations for the magnetic DA  WD ZTF J1901+1458, that was first observed by \citet{Ilaria_moon}. They find it to be significantly less massive and colder than the initial determinations, with $\rm T_{eff} =28\,060 \pm20 \, \rm K$ and radius $ \rm R= 2630 \pm10\, \rm km$, both from SED fitting. Comparing with the WD cooling models of \citet{Althaus2022}, they estimate a mass of $1.30 \pm 0.01 \, \rm M_\odot$ and a cooling age of $460 \,- 490 \rm \,\, Myr$, depending on whether the star has an ONe or CO core.
From our computations, we find that for this same temperature, the radius and cooling age are R = 2406\,km and $\rm t_{cool}\,=$ 506\,Myr, respectively. The age is in good agreement with the previous determination, while our theoretical radius is smaller, which is expected, as it has no hydrogen layer.

\subsubsection{Crystallization and Phase Separation }

In addition to the sequence described previously, we calculate two other test sequences in the WD cooling track to investigate the effect of phase separation at such a high stellar mass. One in which we do not account for phase separation (PS), and the other in which we revert \verb|skye_min_gamma_for_solid| and \verb|skye_max_gamma_for_liq| to their default values in MESA (170 and 200), which decreases the regime treated by Skye's one-component plasma fits, when compared to our baseline model. Changing these limits can create a 10\% difference in the Coulomb parameter $\left< \Gamma \right>$ at crystallization \citep{Skye}, therefore affecting not only the crystallization timeline but also the cooling time. 

We find that, for this stellar mass, the cooling delay introduced by phase separation is minimal. Compared to the model without phase separation, our baseline model is only 16\,Myr older at the time we cease computations, when logL falls below $10^{-4}\,\rm L_\odot$. For the sequence in which we change the solid-liquid regime, the difference in cooling time is even smaller, only 9\,Myr younger than our baseline model.  

As for the effects on crystallization, we present in table \ref{tab:cryst} the cooling times, $T_\mathrm{eff}$, and $\Gamma$ for the three sequences at the start of crystallization ($\phi=0.5$) and at the time the center fully solidifies ($\phi=1$).

\begingroup
\renewcommand{\arraystretch}{1.1}
\begin{table}
    \centering
    \begin{tabular}{l|lcc|lcc} 
    \toprule
     & \multicolumn{3}{c}{$\phi=0.5$}&\multicolumn{3}{c}{$\phi=1$} \\
     \cmidrule(lr){2-4}\cmidrule(lr){5-7}
     %\midrule
      run & $\Gamma$ & $T_\mathrm{eff}$\,(K)  & $t_\mathrm{cool}$ & $\Gamma$ & $T_\mathrm{eff}$\,(K) & $t_\mathrm{cool}$ \\
     \midrule
     baseline & 195& 48530 &81.6& 248& 42840 &137.8 \\%\hline
     noPS &  195& 48470 &79.6& 253 & 41090 & 160.7 \\
     def & 191 & 47900& 81.6& 233& 44220 &118.2\\  
     \bottomrule
    \end{tabular}
    \caption{Coulomb parameter $\Gamma$, Teff, and cooling age ($\rm t_{cool}$) in Myr for our WD sequences at the start ($\phi=0.5$) and full ($\phi=1$) crystallization of the center. Baseline is the sequence we described in \ref{subsec:wd_stage}, noPS is the sequence where we do not account for phase separation, and def refers to a sequence run with Skye's default regime for one-component plasma fits ($\Gamma_{min}=170$, $\Gamma_{max}= 200$).}
    \label{tab:cryst}
\end{table}
\endgroup

At the onset of crystallization, the baseline and no PS sequences agree well in $T_\mathrm{eff}$ and $\Gamma$; however, there is already a 2\,Myr difference in age.  For the sequence with a different solid-liquid regime, the age at the start of crystallization changes only slightly; however, crystallization begins at a smaller $\Gamma$, resulting in a small difference in effective temperatures. These differences increase when comparing the sequences once the center is fully crystallized, with the age spread being larger than at $10\,000$\,K.  We note that the different prescriptions also lead to slight differences in the abundance profiles, which we present in Appendix \ref{append:B}.

\subsubsection{Comparison with previous works}
Here we present a brief comparison of UMWD models calculated by \citet{lauffer} and \citet{camisassa+19} with the one computed in this work. We chose these works because they feature UMWD models that include crystallization temperatures, cooling times, and composition. %\textcolor{magenta}{ADD 1.31 O/Ne althaus2022? composição é da camisassa, cooling times ele tem (de 80000K a 10000K são 1.87Gyr, Teff crist else incluem como ponto no grafico em log, mas ta entre 40 - 50 mil K}

\citet{lauffer} calculated the full evolution for massive and ultramassive white dwarf progenitors, using a high mass-loss scheme from the start of the AGB to skip thermal pulsing calculations, to reach the post-AGB stage and the white dwarf cooling curve. They artificially add He at the beginning of the WD cooling track, since their most massive models lost part of the He-layer due to winds, and account for crystallization and latent heat release, which occur between $\Gamma=215-220$.  Their most massive sequence is a 1.307\,$\rm M_\odot$ WD with a Ne/O/Mg core, which is the result of adopting a smaller nuclear reaction network. It begins crystallization at $\rm T_\mathrm{eff} = 45\,137$~K, when $\rm t_{cool}$\footnote{\citet{lauffer}defines the start of the cooling curve at logL=0}$= 99$\,Myr, and has a final cooling age of 1.31\,Gyr, at $\rm T_\mathrm{eff}=10\,000~K$. 

\citet{camisassa+19} constructed and evolved UMWD using the final models of \citet{siess2010}, which calculated their models until late stages of the TP-AGB. They account for crystallization and phase separation, adopting the ONe phase diagrams of \cite{ps_camisassa}. Their more massive model, which has a mass of $1.29\,\rm M_\odot$, has $t_\mathrm{cool}$\footnote{No precise definition of $\rm t_\mathrm{cool}$=0 is provided}$=2.05$\,Gyr at $\log L=-4$. They present effective temperatures at crystallization only for their H-rich models, which is $\sim38\,000$~K for the 1.29\,$\rm M_\odot$ sequence. 

Despite differences in composition and crystallization parameter $\Gamma$, our model without PS shows a very similar $\rm T_\mathrm{eff}$ ($\Delta \rm T_\mathrm{eff} < 7\%$) to that of \citet{lauffer} upon crystallization, and cooling ages that differ by $\sim 20\%$. Regarding the comparison with the model from \citet{camisassa+19}, we observe cooling times that differ by 0.38 Gyr and a crystallization $\Delta \rm T_\mathrm{eff} $ of at most $\sim$10,000~K. This difference is likely explained by the slight difference in mass and the not-so-small difference in composition. Their models have a $^{20}$Ne to $^{16}$O ratio of $\sim0.5$ and appear to be more carbon-rich than ours, which has $^{20}$Ne/$^{16}$O$\sim0.8$. This could be partially explained by the fact that our sequence starts C-burning with a higher $\rm M_\mathrm{CO}$ than the one from \cite{siess2010} and the shorter time ours stays in the TP-AGB. However, the difference in nuclear networks and nuclear rates cannot be excluded as a possible contributor. 

 We find our work to be in good agreement with both the Newtonian and GR $1.31 \rm M_\odot$ sequence from \citet{Althaus2022}, which were constructed with the same composition as the $1.29 \rm M_\odot$ H-rich model of \citet{camisassa+19}. At $\log \rm L$ = -4, their Newtonian sequence has a cooling age\footnote{\citet{Althaus2022} cooling sequences start at $\rm L \sim \rm L_\odot $, corresponding to $T_\mathrm{eff}\approx 80 \, 000$\,K for their 1.31$\rm M_\odot$ sequences} of 1.89 Gyr, and the inclusion of GR decreases this time in only $\sim0.1$\,Gyr. In both sequences, crystallization starts between $\rm T_\mathrm{eff}$ 50\,000 K and 40\,000 K.

\section{Conclusions}\label{sec:conc}

In this work, we have calculated, using MESA r24.08.1, the entire evolution of a 9\,$\rm M_\odot$ star with Z=0.02. We detailed its evolution from the EAGB to the WD cooling track, conducting comparisons with alternative models to study the impacts of temporal and spatial resolution, thermal pulse suppression, and phase separation. We report core masses at relevant stages of its evolution and present detailed chemical profiles at the end of the EAGB and TP-AGB, as well as at different temperatures on the WD track.

As a result of our calculations, we find a hydrogen-deficient white dwarf of mass 1.313$\,\rm M_\odot$ and He content of $1.5 \times 10^{-5}\,\rm M_\odot$. The interior is 47.7\% $^{16}$O, 39.7\% $^{20}$Ne, 4.2\% $^{24}$Mg, and 3.3\% $^{23}$Na in respect to the total mass, making it an O/Ne/Mg WD. The total mass fractions of $^{12}$C and $^{22}$Ne are 0.386\% and 0.027\%, respectively. The inclusion of ONe phase separation introduces a cooling delay of only $\sim 16$\,Myr.

We find that at this mass, the TP-SAGB stage has only a minor effect on the final mass and composition. As such, calculating only the first pulses or altogether skipping thermal pulses, by adopting a high mass-loss rate \textbf{after} the end of the second dredge-up, is a computationally efficient method to create UMWD models from single evolution, with the benefit of reliable chemical compositions. 

This is, to our knowledge, the most massive white dwarf model from full evolution for which cooling times and detailed abundance profiles are available in the literature. It is also the only near-Chandrasekhar WD model for which both the TP-SAGB and post-AGB were calculated. 

%\subsection{Figures and tables}

%Figures and tables should be placed at logical positions in the text. Don't
%worry about the exact layout, which will be handled by the publishers.

%Figures are referred to as e.g. Fig.~\ref{fig:example_figure}, and tables as
%e.g. Table~\ref{tab:example_table}.

%\begin{figure}
	%\includegraphics[width=\columnwidth]{example}
    %\caption{This is an example figure. Captions appear below each figure.
	%Give enough detail for the reader to understand what they're looking at,
	%but leave detailed discussion to the main body of %the text.}
    %\label{fig:example_figure}
%\end{figure}

%\begin{table}
	%\centering
	%\caption{This is an example table. Captions appear above each table.
	%Remember to define the quantities, symbols and units used.}
	%\label{tab:example_table}
	%\begin{tabular}{lccr} % four columns, alignment for each
		%\hline
		%A & B & C & D\\
		%\hline
		%1 & 2 & 3 & 4\\
		%2 & 4 & 6 & 8\\
		%3 & 5 & 7 & 9\\
		%\hline
%	\end{tabular}
%\end{table}

%\section{Conclusions}

%The last numbered section should briefly summarise what has been done, and describe
%the final conclusions which the authors draw from their work.

\section*{Acknowledgements}
We thank the anonymous referee for their thorough review. \par 
This work was carried out with the financial support of the Conselho
Nacional de Desenvolvimento Científico e Tecnológico (CNPq-Brazil). 
This research has extensively utilized NASA's Astrophysics Data System Bibliographic Service (ADS).
%%%%%%%%%%%%%%%%%%%%%%%%%%%%%%%%%%%%%%%%%%%%%%%%%%
\section*{Data Availability}

Following MESA guidelines, the inlists and modified rse files used to perform our calculations can be found in the Zenodo MESA repository at \url{https://doi.org/10.5281/zenodo.18868476} 
Cooling tables for the WD cooling tracks will be available at CDS via \url{https://cdsarc.cds.unistra.fr/viz-bin/cat/J/MNRAS}, once the paper is published.



\bibliographystyle{mnras}
\bibliography{cite} % if your bibtex file is called example.bib

%\begin{thebibliography}{99}
%\end{thebibliography}

%%%%%%%%%%%%%%%%%%%%%%%%%%%%%%%%%%%%%%%%%%%%%%%%%%

%%%%%%%%%%%%%%%%% APPENDICES %%%%%%%%%%%%%%%%%%%%%

\appendix

\section{Additional WD profiles} \label{append:B}

%We present in Table~\ref{tab:wd_extra} some physical quantities of observational interest at different temperatures. 
%\begin{table}
 %   \centering
 %   \begin{tabular}{|c|c|c|c|c|} \hline
    
  %   $T_\mathrm{eff}$ (K) & $\mathrm{t_{cool}}\, \mathrm{(yr)} $& Log L & Log g & R (km) \\\hline
  %  292\,617 & 5.00 $\times 10^1$ & 1.99 & 9.38& 2\,684 \\ 
  %   %\hline
   % 150\,571 & $2.80 \times 10^4$ & 0.80 & 9.42 & 2\,569 \\
   % % \hline 
   % 100\,000 & $1.07  \times 10^5$& 0.07 & 9.43 & 2\,527 \\
    % \hline 
  % 70\,000 &  $9.06  \times 10^6$ & -0.56 &  9.45 & 2\,468 \\
    % \hline
  %  40\,000 & $1.81  \times 10^8$ & -1.55 &  9.47 & 2\,420 \\
    % \hline 
  %  20\,000 & $9.09  \times 10^8$ & -2.77 &  9.48 & 2\,400 \\
    % \hline
  %  10\,000 & $1.66  \times 10^9$ & -3.96 &  9.48 & 2\,396 \\
  %   \hline
  %  \end{tabular}
  %  \caption{ }
  %  \label{tab:wd_extra}
%\%end{table}

\begin{figure*}
    \centering 
    \begin{subfigure}{0.48\textwidth}
        \raggedright
        \includegraphics[width=\linewidth]{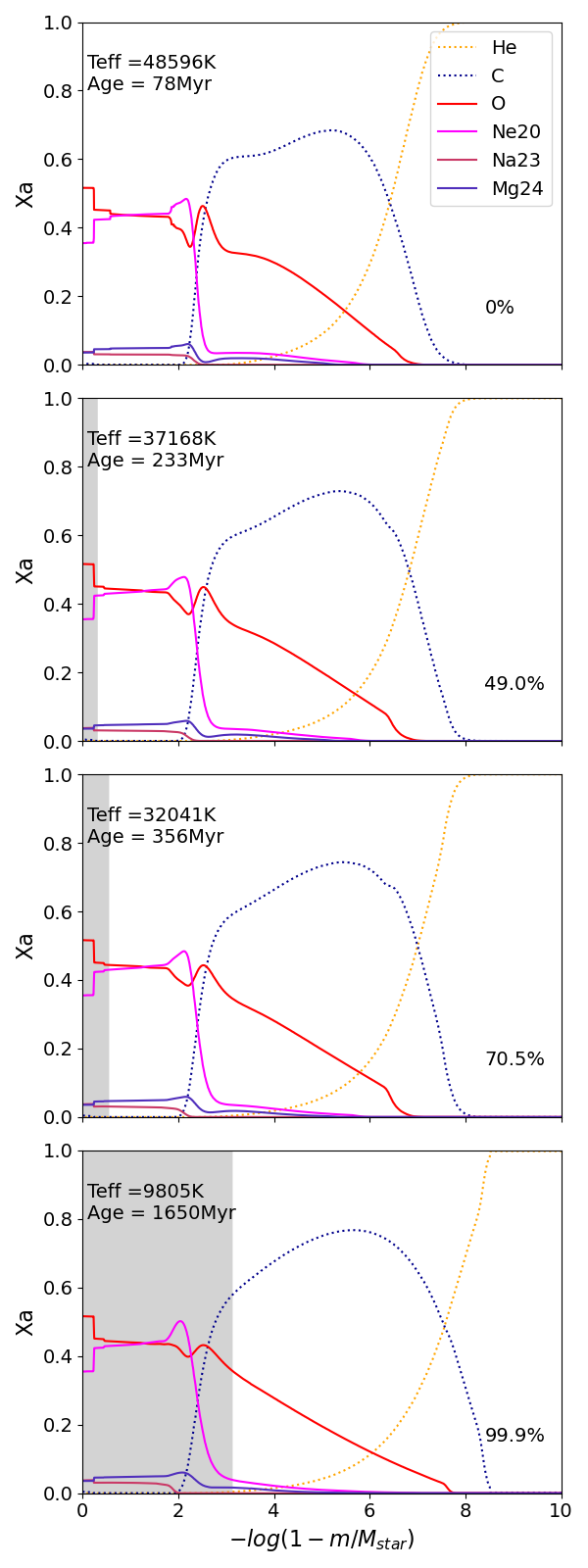}
    \end{subfigure}
    \begin{subfigure}{0.48\textwidth}
        \raggedleft
        \includegraphics[width=\linewidth]{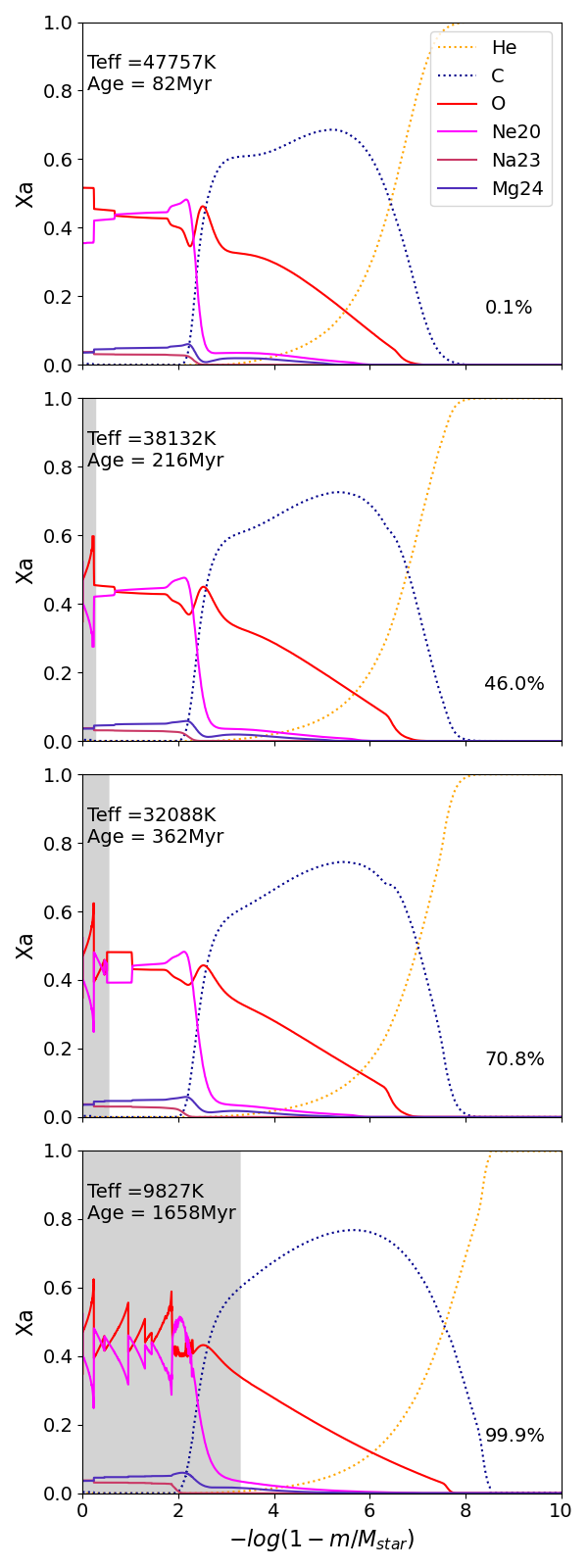}
    \end{subfigure}
    \caption{Abundance profiles for our additional $1.313\,\rm M_\odot$ WD sequences  at four different moments in time. For clarity, we once again include in the plots only the most relevant species, and represent crystallization in grey. Panels on the left show the profiles for the sequence in which we do not include phase separation, while panels on the right show it for a sequence with the default solid-liquid regime of Skye.}
    \label{fig:wd_profile_NoPS}
\end{figure*}

As mentioned in Subsection~\ref{subsec:wd_results}, both the inclusion of phase separation and the adoption of a different liquid-solid regime result in slight changes to the abundance profiles. Thus, we present in Figure~\ref{fig:wd_profile_NoPS} profiles at various temperatures for our additional WD models. 
On the left are the profiles for the sequence calculated without Phase Separation (noPS). On the right, we show profiles for the sequence that accounts for phase separation with Skye's default liquid-solid regime (def). We choose our profiles so that the "noPS" case has similar crystallized fractions to the ones shown in Figure~\ref{fig:wd_profile}, and "def" the closest $T_\mathrm{eff}$ to "noPS".

%%%%%%%%%%%%%%%%%%%%%%%%%%%%%%%%%%%%%%%%%%%%%%%%%%

% Don't change these lines
\bsp	% typesetting comment
\label{lastpage}
\end{document}